\documentclass{cs20proc}

\usepackage{kantlipsum}

\editors{S. J. Wolk}
\publisher{Zenodo}
\conference{The 20th Cambridge Workshop on Cool Stars, Stellar Systems, and the Sun}
\conferencedate{2018}

\title{Meter- to Millimeter Emission from Cool Stellar Systems:
Latest Results, Synergies Across the Spectrum, and Outlook for the Next Decade}
\author{
Jan Forbrich,$^{1,2}$ 
Peter K.~G. Williams,$^{2}$
Emily Drabek-Maunder,$^{3,4}$
Ward Howard,$^{5}$
Moira Jardine,$^{6}$
Lynn Matthews,$^{7}$
Sofia Moschou,$^{2}$
Robert Mutel,$^{8}$
Luis Quiroga-Nu\~nez,$^{9}$
Joseph Rodriguez,$^{2}$
Jackie Villadsen,$^{10}$
Andrew Zic,$^{11}$
Rachel Osten,$^{12}$
Edo Berger,$^{2}$ \&
Manuel G\"udel$^{13}$
}

\affiliation{
$^{1}$ Centre for Astrophysics Research, University of Hertfordshire, College Lane, Hatfield AL10 9AB, UK\\
$^{2}$ Harvard-Smithsonian Center for Astrophysics, 60 Garden Street, Cambridge, MA 02138, USA\\
$^{3}$ School of Physics and Astronomy, Cardiff University, The Parade, Cardiff, CF24 3AA, UK\\
$^{4}$ Royal Observatory Greenwich, Blackheath Ave, London SE10 8XJ, UK\\
$^{5}$ Department of Physics and Astronomy, University of North Carolina at Chapel Hill, Chapel Hill, NC 27599-3255, USA\\
$^{6}$ Centre for Exoplanet Science, SUPA, School of Physics and Astronomy, University of St Andrews, North Haugh, St Andrews KY16 9SS, UK\\
$^{7}$ MIT Haystack Observatory, 99 Millstone Road, Westford, MA 01886, USA\\
$^{8}$ Department of Physics and Astronomy, University of Iowa, 203 Van Allen Hall, Iowa City, IA, 52242, USA\\
$^{9}$ Leiden Observatory, Leiden University, P.O. Box 9513, 2300 RA Leiden, The Netherlands\\
$^{10}$ National Radio Astronomy Observatory, 520 Edgemont Road, Charlottesville, VA 22903, USA\\
$^{11}$ Sydney Institute for Astronomy (SIfA), School of Physics, The University of Sydney, NSW 2006, Australia\\
$^{12}$ Space Telescope Science Institute, 3700 San Martin Drive, Baltimore, MD 21218, USA\\
$^{13}$ University of Vienna, Department of Astrophysics, T\"urkenschanzstrasse 17, 1180 Vienna, Austria\\
}

\shorttitle{Radio emission from cool stars}
\shortauthors{Forbrich et al.}

\abs{Radio observations of cool stellar systems provide unique information on their magnetic fields, high-energy processes, and chemistry. Buoyed by powerful new instruments (e.g. ALMA, JVLA, LOFAR), advances in related fields (e.g., the {\it Gaia} astrometric revolution), and above all a renewed interest in the relevant stellar astrophysics, stellar radio astronomy is experiencing a renaissance. In this splinter session, participants took stock of the present state of stellar radio astronomy to chart a course for the field's future.}

\begin{document}

\maketitle

\section{Introduction}

In this splinter session, participants discussed stellar radio astronomy with an emphasis on latest results, synergies across the spectrum, and an outlook for the next decade. A total of ten presentations was selected to complement radio-themed plenary talks by Rachel Osten, Meredith MacGregor, and Jan Forbrich, published elsewhere in these proceedings. Additionally, poster presenters were able to highlight their work in a poster pops slot. The session concluded with an open discussion.

This splinter session mainly focused on nonthermal radio emission. In the context of cool star radio astronomy, this traditionally encompassed mainly (gyro)synchrotron radiation at centimeter and millimeter wavelengths, i.e., emission from (mildly) relativistic electrons in magnetic fields. But recent years have shown a variety of other ways in which radio observations can deliver astrophysical insight. The discovery of brown dwarf radio emission has rekindled interest in electron cyclotron maser (ECM) emission as just one of several intriguing physical mechanisms in this regime, many of which are stepping stones toward studies of low-frequency exoplanetary radio emission. VLBI observations of stellar radio emission can be used to obtain precision astrometry -- and hence precise measurements of fundamental parameters -- of ultracool dwarfs or young stellar objects that are not accessible even to {\it Gaia}. Broadband, high-time-resolution studies of flares reveal the detailed plasma processes operating in the coronae of other stars.

The world's best present-day radio observatories -- chiefly the VLA and ALMA -- launched the stellar radio astronomy renaissance. This is particularly true for constraints on nonthermal emission, which are particularly dependent on continuum sensitivity. Spectral indices, polarization, and variability all require high S/N detections to derive meaningful constraints. The upgraded VLA and VLBA as well as Arecibo now provide such information in the cm wavelength range, ALMA is doing so in the mm wavelength range, and LOFAR, MWA, LWA, HERA, and GMRT are making new forays into the low-frequency range. As a result of these newly feasible measurements, we are obtaining unique information on the role of magnetic fields on various scales and a new and complementary perspective on high-energy processes. These scales now even include outflows/jets from protostars and T~Tauri stars, providing initial evidence for magnetic fields in young stellar environments.

The new generation of low-frequency radio observatories promises to figure especially prominently in the upcoming decade of stellar radio astronomy. These observatories can monitor the whole sky nearly continuously, surveying for stellar flares, coronal mass ejections, and other heliospheric phenomena. They are also expected to be the only facilities that will be able to directly probe the magnetic fields of extrasolar planets, through auroral radio bursts, providing unique insight into an essential facet of exoplanetary habitability.

There are important synergies of nonthermal radio emission with other wavelength ranges, as featured in this splinter session. These are mainly due to the beginning of radio time domain astronomy with the advent of sensitive continuum receivers, as seen for instance in the nearly-complete MeerKAT telescope. It is now possible to systematically study high-energy processes in both the X-ray and the radio time domain, and capabilities for doing so will increase dramatically when eROSITA commences operations. Understanding the correlations between optical and radio variability will be essential in the era of first TESS, then LSST.

In the following, we briefly present highlights from the session presentations, in the order of appearance.


\section{Session highlights}

\subsection{Simulating Radio Emission from Low Mass Stars (Moira Jardine)}

One of the enduring puzzles about very low mass stars is that they appear to be radio bright (relative to their X-ray luminosity). Most solar-like stars follow a well-defined relationship between their radio and X-ray luminosities known as the {\it G\"udel-Benz relation}, \citealp{ben94}). This can be understood as two forms of emission (the thermal X-ray emission and the non-thermal radio emission) that originate from the same population of electrons accelerated in flares. The puzzle about the very low-mass stars is that they appear to have excess radio emission over and above what might be expected from this process. 

\begin{figure}
\includegraphics[width=\linewidth]{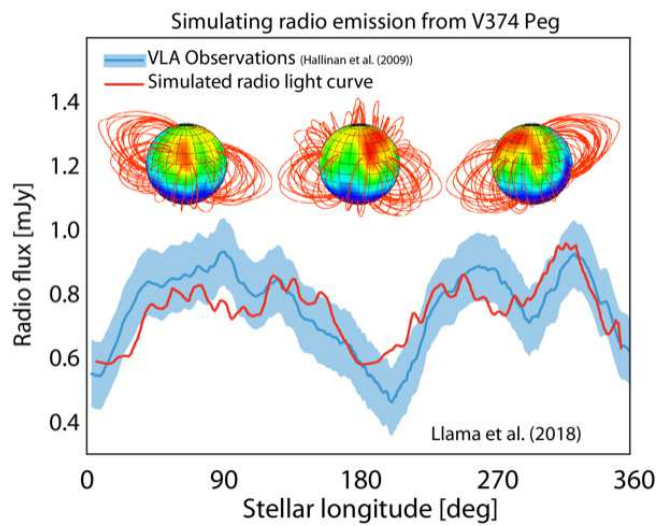}
\caption{Simulating radio emission from V374~Peg, see \citet{lla18}. \label{fig_jardine}}
\end{figure}

One possible explanation is that another process (electron-cyclotron maser emission) is contributing in these objects. To explore this we have taken (almost) simultaneous observations of the surface magnetic field for V374~Peg and its radio lightcurve from the VLA. By extrapolating the 3D stellar magnetic field and modelling the electron cyclotron emission we have produced a synthetic radio lightcurve that fits well to the observed one (see Figure~\ref{fig_jardine}; \citealp{lla18}). This demonstrates the role of the field geometry in determining the nature of the radio emission and provides a new technique for studying the radio emission from low mass stars.

\subsection{Resolved imaging of the quiet and flaring radio corona of active M~dwarfs (Jackie Villadsen)}

Active M dwarfs produce bright radio emission at GHz frequencies: quiescent emission, incoherent flares, and coherent bursts.  The quiescent emission and incoherent flares are generally attributed to gyrosynchrotron emission from mildly relativistic electrons spiraling around magnetic field lines, although the electron cyclotron maser may be responsible for the quiescent emission from V374~Peg \citep{lla18}.  As such, radio emission provides the only direct means of observing energetic particles in stellar atmospheres. Understanding particle acceleration in stellar atmospheres is important to predict the energetic particle flux that can interact chemically with exoplanet atmospheres \citep[e.g.,][]{howard2018}.  However, in modeling the radio light curve and spectral index, the properties of the energetic electron distribution are degenerate with the magnetic field strength and source size.  Measuring the sizes of radio sources, and localizing them relative to the photosphere, thereby anchoring them to photospheric magnetic field measurements, can help relieve this degeneracy.

\begin{figure}
\centering
\includegraphics[width=0.9\linewidth]{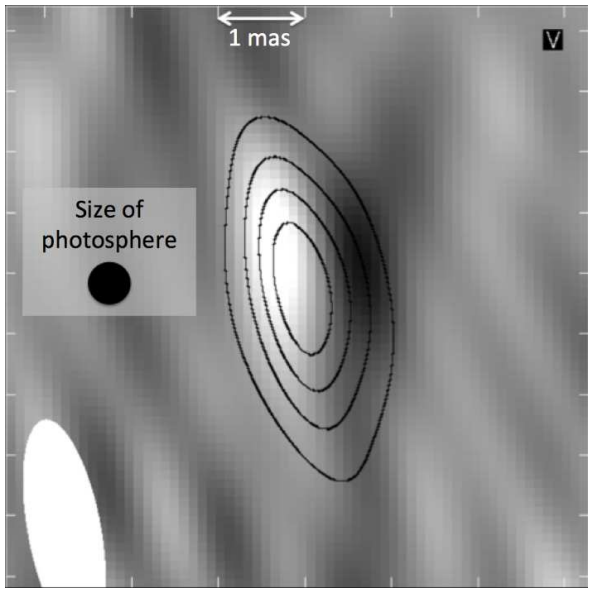}
\caption{VLBA 8.4 GHz image of UV~Cet: Stokes $V$ (grayscale) overlaid with Stokes $I$ (contours). The white oval shows the synthesized beam.  Right polarization (white, originating from a region with northern magnetic polarity) and left polarization (black, originating from a region with southern magnetic polarity) peak at locations separated by roughly a stellar diameter, suggestive of emission associated with the stellar magnetic dipole field.  \label{fig_villadsen}}
\end{figure}

The Very Long Baseline Array (VLBA) at 8.4~GHz achieves a resolution 1-3 times larger than the photospheres of nearby M~dwarfs, enabling source localization and size measurement to a fraction of the photospheric diameter in cases with high signal-to-noise ratio.  VLBI has been used for astrometry of active M~dwarfs at 1.4~GHz \citep{benz1991,benz1995} and 8.4~GHz \citep{bower2009,bower2011}.  In addition, observations at 8.4~GHz have shown that the radio corona of active M~dwarfs is sometimes resolved: \cite{benz1998} observed 2 lobes separated by 4 stellar diameters during a flare on M6 dwarf UV~Ceti, and \cite{pestalozzi2000} observed a radio source 1.7 times larger than the photosphere during quiescent and flaring emission.  2015 8.4-GHz VLBA observations of UV~Cet (Figure~\ref{fig_villadsen}) and AD~Leo, with three 4-hour epochs per target, found distinct behavior on these two stars (Villadsen et al., in prep).  AD~Leo was resolved in 2 of 3 epochs with source size consistent with the photospheric diameter in all epochs, suggestive of quiescent radio emission originating from the low corona, with a minutes-long flare in one epoch offset from the quiescent source by 0.9 stellar diameters. The quiescent emission from UV~Cet was resolved in 3 of 3 epochs, all consistent with a source size of $\sim$3 by 2 stellar diameters, and with each epoch containing a coherent burst associated with a radio aurora.  In two of three epochs, UV~Cet's quiescent emission shows a dipolar structure in Stokes V.  The large scale and dipolar structure of UV~Cet's quiescent emission suggest that the radio emission originates from a population of electrons in the large-scale magnetic dipole field.  Such studies can be carried much further in future if the Next Generation VLA offers long baseline capabilities, which would provide the sensitivity to image thermal emission from nearby stars, including the free-free or gyrosynchrotron emission associated with stellar coronal mass ejections \citep{osten_ngVLA}. 

\subsection{Examining Stellar CME candidates: Does the solar flare-CME relation extend to other stars?\protect\footnote{See poster at https://doi.org/10.5281/zenodo.1405132} (Sofia Moschou)}

A well established relation between solar CMEs and flares has been revealed thanks to
decades of direct observations. Solar statistical studies (e.g., \citealp{yas09,aar11,dra13})
show that strong flares are associated with faster and
more massive CMEs and their correlation increases with increasing respective energies. The
continuously growing number of confirmed extrasolar systems, particularly around M dwarfs,
requires the evaluation of the impact that stellar activity, and in particular CMEs and flares,
might have on habitability. Stellar flares and super-flares ($E_k>10^{33}$~erg) have been observed in
different wavelengths from X-rays to radio wavelengths (e.g., \citealp{ost05,hue10,not16}). 
However, direct imaging of stellar CMEs is currently
impossible, so indirect observational methods need be employed. The three main observational
techniques for capturing stellar CME signatures are measuring a) Type II radio bursts, b)
Doppler shifts in UV/optical lines, and c) continuous absorption in the X-ray spectrum. Here we
examine the most probable CME candidates up to date together with their kinematics and
energetics (Moschou et al. 2018, {\it in prep.}, from now on M18).

\begin{figure}
\centering
\includegraphics[width=0.89\linewidth]{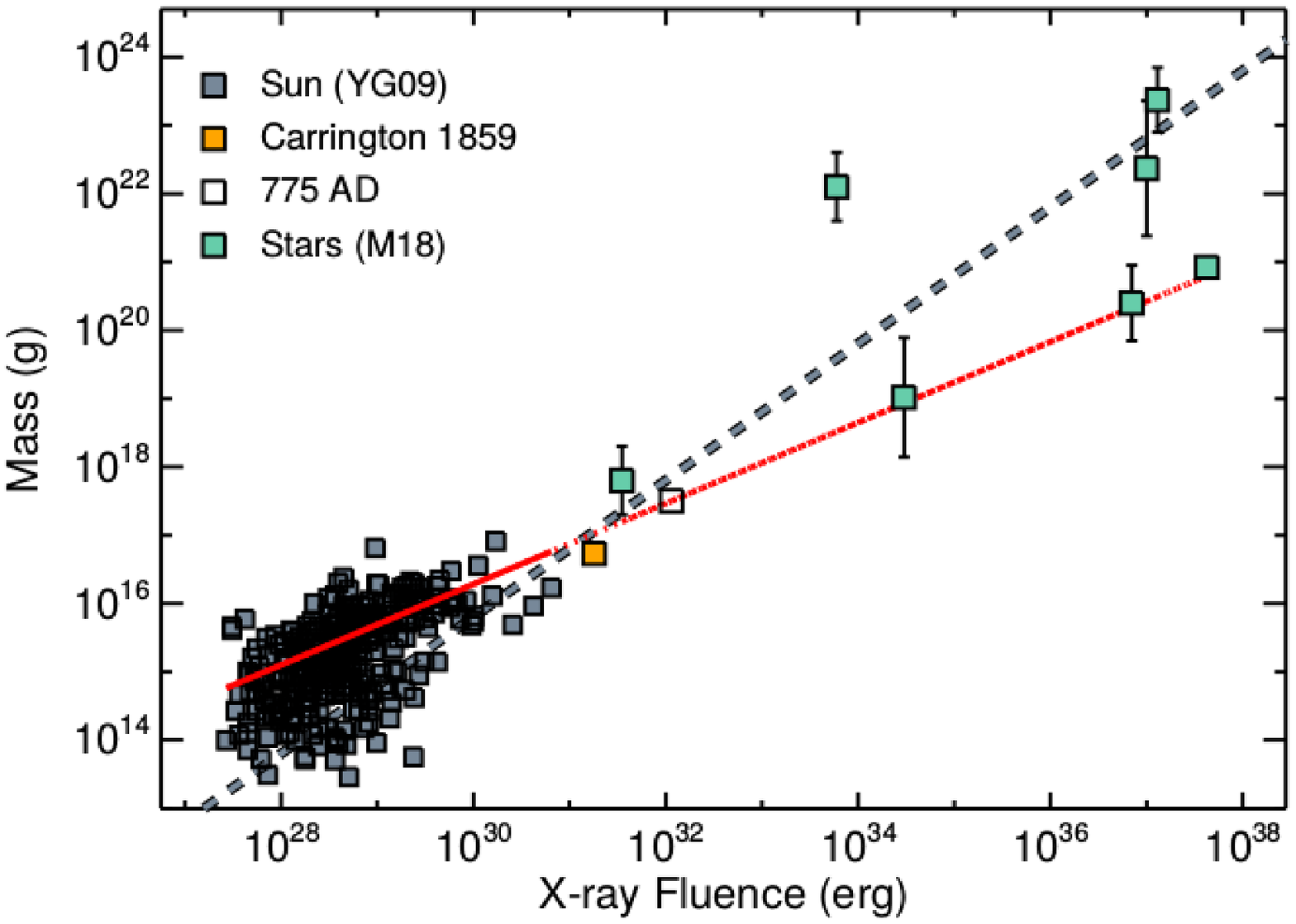}
\includegraphics[width=0.89\linewidth]{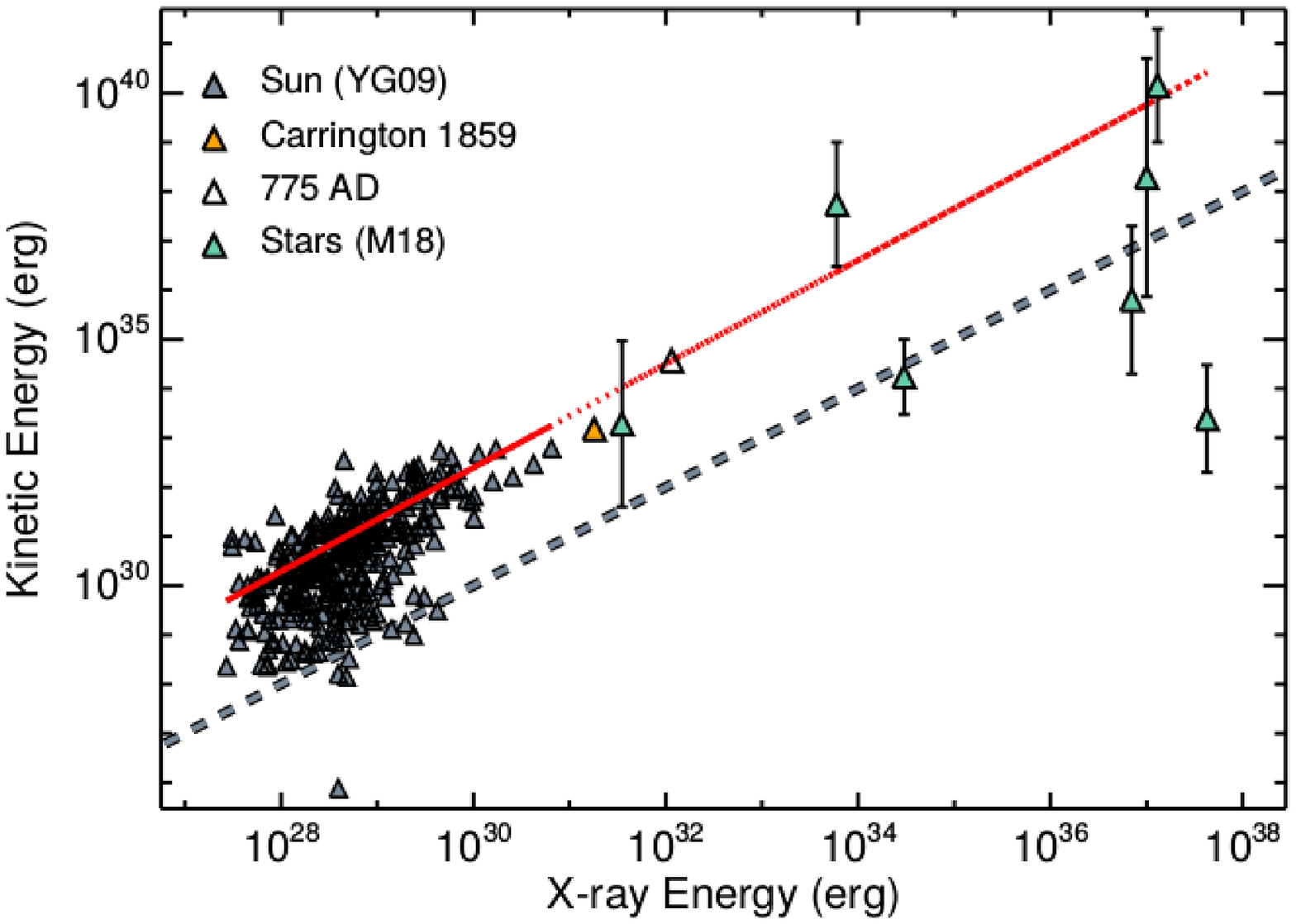}
\caption{CME mass (upper panel) and kinetic energy (lower panel) with respect to the X-ray flaring energy of the
associated CME. Solar events are indicated with filled black symbols, while stellar events are indicated
with green filled symbols. Historic energetic solar events are also shown (orange and white symbols). Our
preliminary results compiled using historic observations indicate that in the stellar regime the solar trends
stand with the CME kinetic energy reaching a plateau gradually. \label{fig_moschou}}
\end{figure}

In an earlier paper \citet{mos17} analyzed an extreme super-flare from the Demon star
(Algol binary) using the CME cone model, a geometric model often used in solar physics.
\citet{mos17} estimated the CME mass and kinetic energy which were in agreement
with the solar trend, given the spread of the solar events and the large errors they calculated.
The super-flare was observed in X-rays through continuous absorption \citep{fav99}
 and is probably the best CME candidate to date. Measurements of Doppler shifts in lines such as H$_\alpha$ \& H$_\gamma$ can also be used to estimate stellar outflows (e.g., 
\citealp{vid16}). There are a few historic measurements that captured speeds larger than the local
escape speeds. Finally, Type II radio bursts are associated with CME shock waves in the Sun.
Unfortunately, there is no Type II radio burst observed in another star yet (e.g., \citealp{cro18}). 
We use all the X-ray continuous absorption and EUV blueshift observations consistent
with an escaping mass up to date in an effort to examine the extension of the solar CME flare
relation towards the high-energy limit for active stars (M18). Our preliminary results indicate that
a similar energy partition between flares and CMEs stands for active stars with a CME kinetic
energy that gradually reaches a plateau with increasing flaring X-ray energy (M18, see 
Figure~\ref{fig_moschou}).

\subsection{Evryscope detection of the first Proxima superflare: Habitability Impacts and Multi-Messenger Opportunities (Ward Howard)}

\begin{figure}
\includegraphics[width=\linewidth]{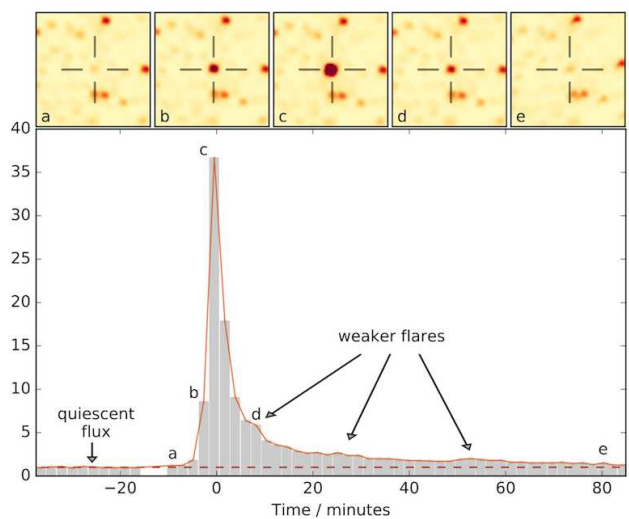}
\caption{Evryscope discovery of a naked-eye-brightness superflare from Proxima in March 2016. The y-axis is the flux increase over Proxima's median g-band flux from the previous hour. Bars show the integration time of each individual flux measurement. Insets display cutout images over the course of the flare. \label{fig_howard}}
\end{figure}

In March 2016, the Evryscope observed the first naked-eye-visible superflare from Proxima Centauri (see Figure~\ref{fig_howard}). The Evryscope array of small optical telescopes recorded the superflare as part of an ongoing survey of all bright southern stars, monitored simultaneously at 2 minute cadence since 2015. In light of large Proxima flares observed in the X-ray \citep{gue04} and sub-mm \citep{mac18}, we explore the contributions that continuous Evryscope monitoring bring to multi-wavelength studies of M-dwarf flare morphology and evolution, and the impacts on space weather environments of planets such as Proxima b. By modeling the photochemical effects of particle events accompanying large flares in a recently-accepted letter, we find repeated flaring is sufficient to reduce the ozone column of an Earth-like atmosphere at the orbit of Proxima b by 90\% within five years and 99.9\% over geologically-short timescales. Assuming complete ozone loss, surface UV-C levels during the Evryscope superflare reach ~100$\times$ the intensity required to kill simple UV-hardy microorganisms, suggesting that life would struggle to survive in the areas of Proxima b exposed to these flares.

With the recent launch of TESS \citep{ric14}, which observes most stars for 27 days, multi-year Evryscope monitoring will help constrain both the emission mechanisms and habitability impacts of stellar flares for all the bright stars in the TESS field. In conjunction with Evryscope's long-wavelength radio counterpart, we discuss how simultaneous monitoring of the visible sky by Evryscope and the LWA \citep{hal17} may localize radio flares and potentially exoplanet auroral radio bursts. Furthermore, a cycle 6 ALMA proposal (PI: M. MacGregor) for 40 hours of simultaneous Evryscope and ALMA observations of large sub-mm flares on Proxima has been accepted with a priority grade of A. We discuss how we may constrain any associated flare blackbody emission to sub-mm events with Evryscope, providing insight into the currently-unknown emission mechanism and habitability impacts of these flares.

\subsection{Low-frequency GMRT observations of ultra-cool dwarfs\protect\footnote{See talk slides at https://doi.org/10.5281/zenodo.1407817} (Andrew Zic)}

Ultra-cool dwarfs (UCDs; spectral type $>$M7) are objects which span the gap between the lowest mass stars and the highest mass planets. Their fully-convective interiors initially led beliefs that they should not have large-scale magnetic fields \citep{dur93}. In addition, standard stellar magnetic activity tracers such as H$\alpha$ and X-ray luminosity drop sharply for UCDs \citep{moh03}. The strong non-thermal radio emission detected from around 10\% of UCDs was therefore unexpected, as it shows that these objects are capable of generating strong, large-scale magnetic fields. Due to their very low H$\alpha$ and X-ray luminosities, radio-frequency observations have been vital to understanding the physical processes operating in UCD magnetospheres. Most radio-frequency studies of UCDs have taken place between 4 and 8 GHz, where traditional radio interferometers are typically most sensitive. Hence, the nature of low-frequency emission ($<$1.4~GHz) from UCDs remains relatively unexplored, and could provide insights into the coronal and magnetospheric properties of these stars. In particular, low-frequency observations probe the optically-thick side of the quiescent UCD spectral energy distribution (SED). This allows the spectral turnover frequency to be accurately determined by comparing low-frequency (optically-thick) and high-frequency (optically-thin) flux density measurements. This provides a further constraint on coronal properties of the UCD that can be obtained by modeling their non-thermal gyrosynchrotron radiation. 

We have observed nine UCDs at around 610 and 1300 MHz with the Giant Metrewave Radio Telescope. We have detected quiescent gyrosynchrotron radiation from two UCDs: LSPM J1314+1320, and 2MASS J0746+20, making these the lowest-frequency detections of UCDs to date. Using these new low-frequency measurements, and measurements at higher frequencies available in the literature, we are able to determine that the spectral turnover of LSPM J1314+1320 is at approximately 2.2 GHz. The nonsimultaneous SED for LSPM J1314+1320 is shown in the attached figure. Using this information, we make order-of-magnitude estimates of the coronal magnetic field strength of 20 G, and electron number density $10^7$~cm$^{-3}$. One important limitation is that our low-frequency measurements were non-simultaneous with measurements at higher frequencies, which introduces uncertainty into the radio spectral shape and therefore to the coronal parameter estimates. We plan to take simultaneous multi-frequency observations in the near future to overcome this issue.

\begin{figure}
\includegraphics[width=\linewidth]{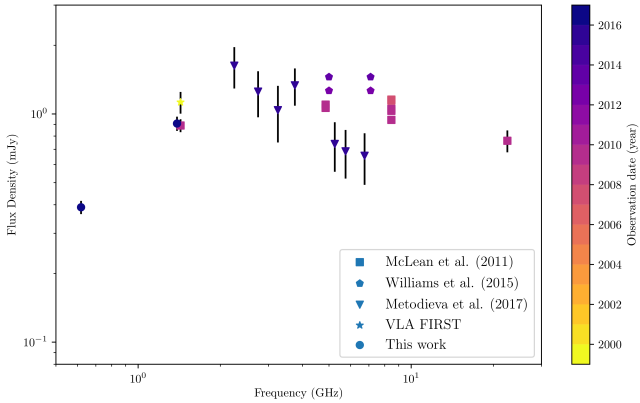}
\caption{Non-simultaneous SED of LSPM J1314+1320, with measurements taken from the Very Large Array FIRST survey, \citet{mcl11}, \citet{wil15}, \citet{met17}, and Zic et al. (this work; {\it in prep.}). Despite long-term variability evident at higher frequencies, a turnover at approximately 2.2~GHz is apparent. \label{fig_zic}}
\end{figure}

\subsection{Serendipitous discovery of variability at low radio frequencies: the case of M-dwarf star Ross 867 (Luis Quiroga-Nu\~nez)}

Dynamic activity in the stellar atmosphere of late-type dwarf stars with a spectral type of $\sim$M4 has widely been studied before (e.g.,  \citealp{boc07,wes08}. Serendipitously, we have discovered radio emission from the binary system Ross 867-868 while inspecting archival Giant Metrewave Radio Telescope (GMRT) observations nearby the galaxy cluster RXJ1720.1+2638. The binary system consists of two M-dwarf stars separated by around 200~AU that share similar characteristics such as spectral type, emission at optical, infrared and X-rays, and astrometric parameters. 

\begin{figure}
\includegraphics[width=\linewidth]{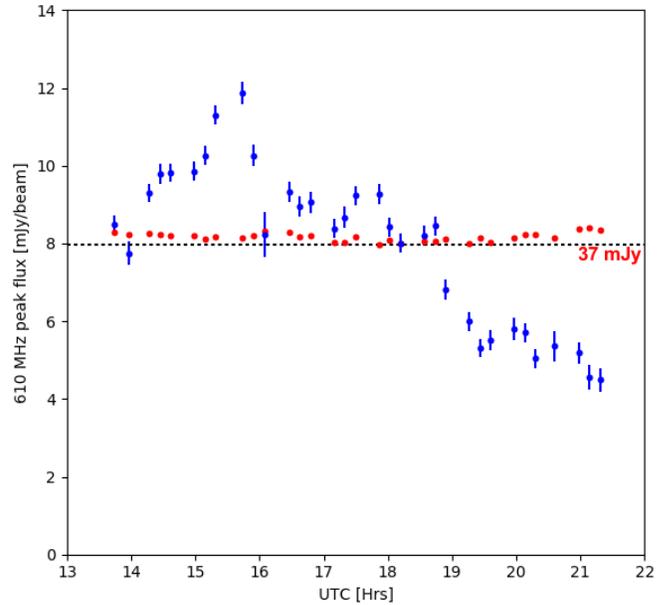}
\caption{Radio light curve for Ross 867 obtained using GMRT data at 610 MHz on 24 July 2011. The blue points represent the Ross 867 emission and the red points show the emission of the close-by quasar J171949+263007, while the horizontal dashed line represents the normalized mean peak flux of both sources. \label{fig_quiroganunez}}
\end{figure}

By searching within various archives, more observations were found of this system, ranging from 1984 with the Very Large Array (VLA) in 1984 to Low Frequency ARray (LOFAR) in 2017. While in most of the observations, Ross 867 has been detected (except for LOFAR at 145 MHz due to dynamic range limitations), Ross 868 remains undetected in the radio archival data inspected (VLA, GMRT and LOFAR). Data from the GMRT at 610 MHz taken on 24 July 2011 shows that the radio emission from Ross 867 is highly variable on hour time scales, with the highest peak flux of 12 mJy/beam. Figure~\ref{fig_quiroganunez} shows the light curve for Ross 867 over more than 8 hours, where the emission of a close-by quasar is shown as reference ($\sim$37~mJy). Further studies will include a radio circular polarization measurement and a comparison of the positions of the optical and radio emissions using Gaia DR2 data and (proposed) EVN VLBI observations.

\subsection{Occultations as a Window into Circumstellar Architecture: RW Aurigae (Joseph Rodriguez)}

\begin{figure}
\centering 
\includegraphics[width=0.6\columnwidth]{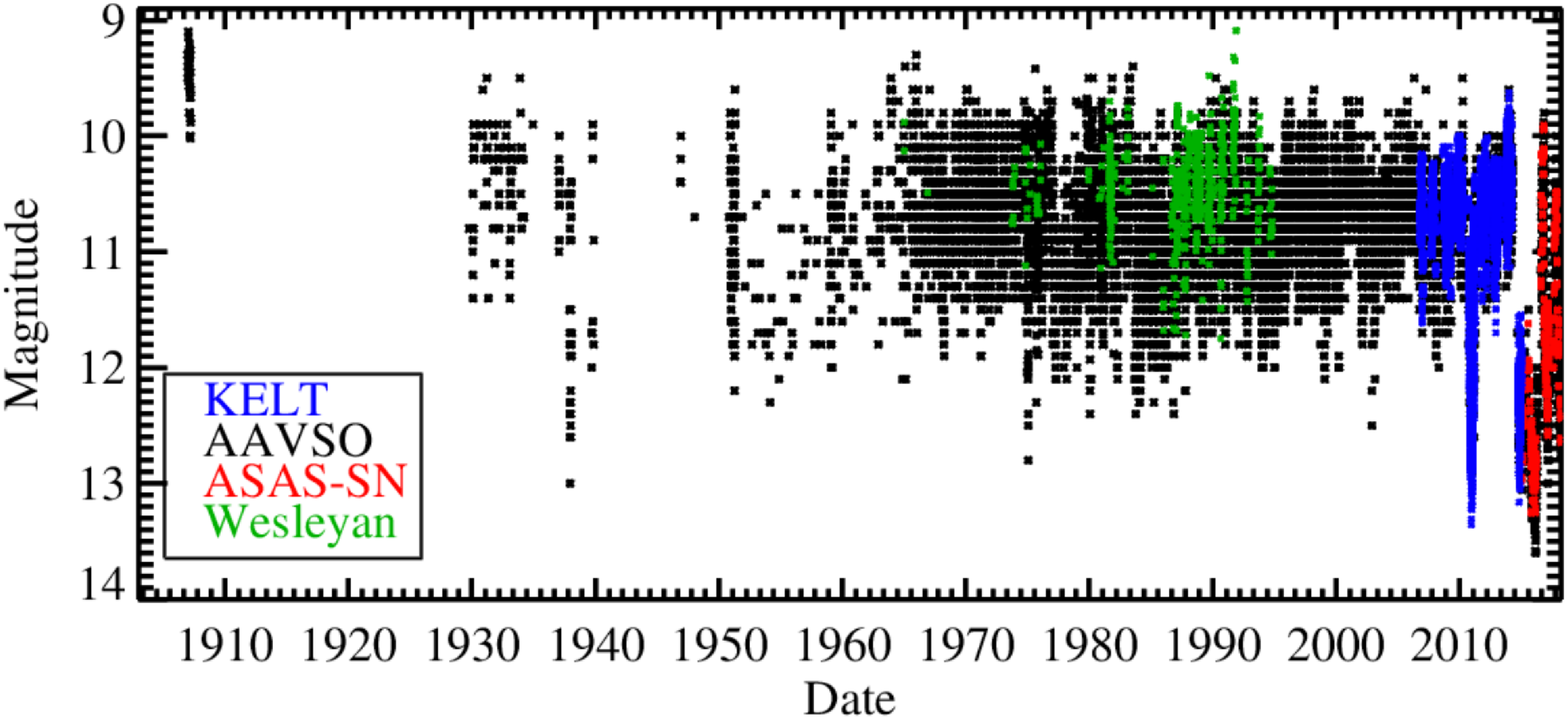}
\includegraphics[width=0.7\columnwidth]{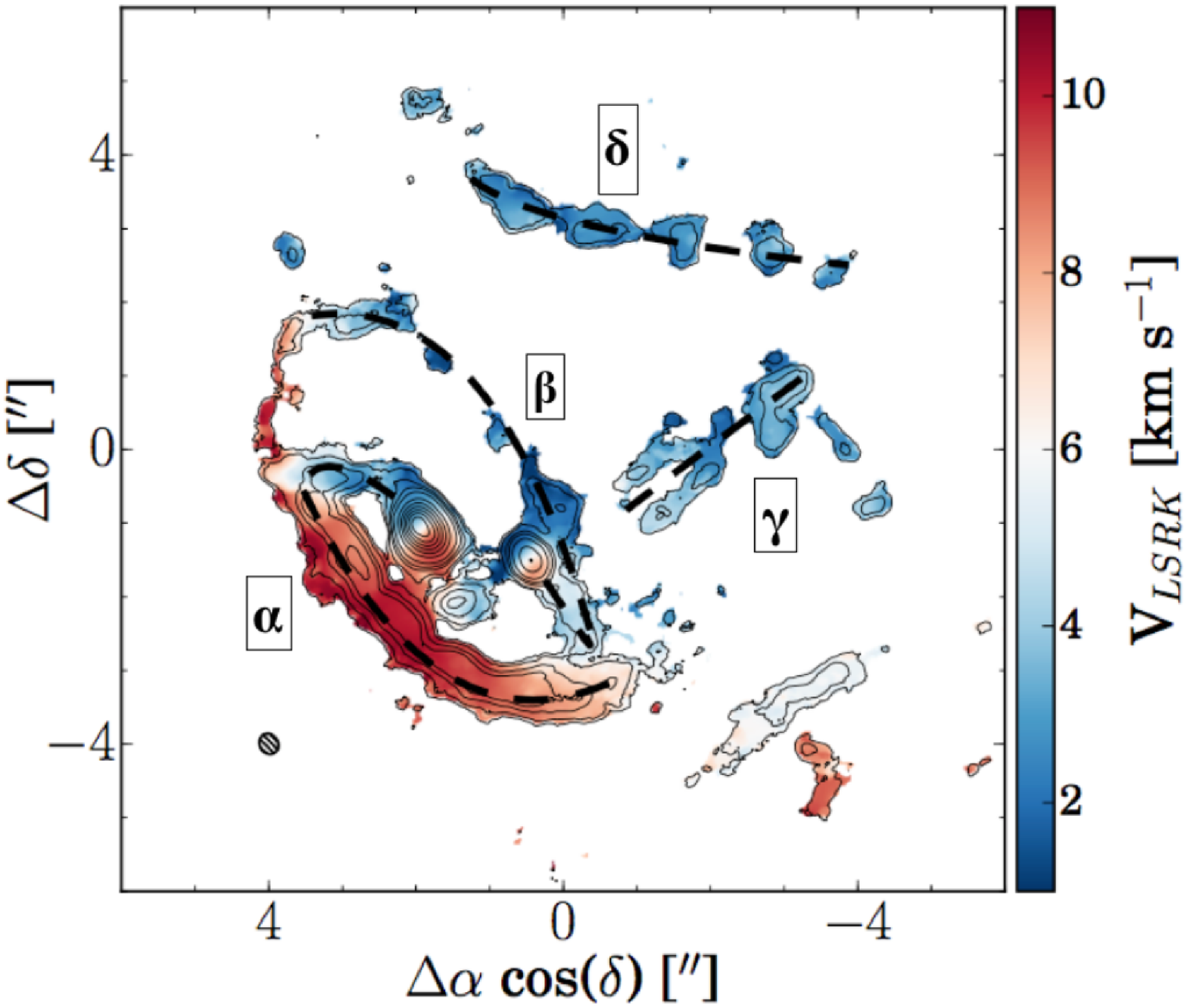}
\caption{(Upper panel) Recreation of Figure 1 from \citet{Rodriguez:2013, Rodriguez:2016} showing the full $\sim$110 year long lightcurve from AAVSO (Black), Wesleyan (Green), KELT (Blue), and ASAS-SN (Red) of RW Aur. (Lower panel) Figure~9 from \citet{Rodriguez:2018} showing the Moment-1 map overlaid with moment-0 contours. The original tidal arm discovered by \citet{Cabrit:2006} is labeled ``$\alpha$", the second counter spiral arm  is labeled ``$\beta$''and the two new tidal streams identified are labeled ``$\delta$" and ``$\gamma$". The dashed black line outlines the two spiral arms that form the apparent ring and the additional tidal streams to the NW. }
\label{fig_rodriguez}
\end{figure}

The circumstellar environment around a young stellar object can directly influence the eventual architecture of planets that will form from the gas and dust within its circumstellar disk. Specifically, the gravitational forces from a stellar companion can significantly disrupt the gas and dust within a circumstellar disk \citep{Clarke:1993}. The archetype for the impact of binary interactions on disk evolution and possibly planet formation is the classical T Tauri star, RW Aurigae. From $^{12}$CO mapping using the Plateau de Bure Interferometer (PdBI), it was discovered that the disk around RW Aur A had recently undergone a stellar fly-by from its companion RW Aur B, resulting in a short truncated disk around RW Aur A and a large tidal stream wrapped around both stars \citep{Cabrit:2006}. Recently, RW Aurigae was re-observed at high spatial resolution ($\sim0.25''$) using the Atacama Large Millimeter/submillimeter Array (ALMA). From these observations, multiple tidal streams were discovered (\citealp{Rodriguez:2018}, see Figure~\ref{fig_rodriguez}). The additional tidal streams suggest that multiple fly-bys of RW Aur B have occurred, repeatedly disrupting the circumstellar disk around RW Aur A.  

Additionally, the RW Aurigae system has been photometrically observed since the late 1890's (\citealp{Beck:2001}, see Figure~\ref{fig_rodriguez}). In 2010, RW Aurigae experienced a $\sim$2 magnitude dimming event for $\sim$180 days \citep{Rodriguez:2013} with no event of similar depth and duration ever being seen prior \citep{Berdnikov:2017, Rodriguez:2018}. Since the 2010 event, RW Aurigae has undergone four additional dimming events, each varying in depth and duration \citep{Petrov:2015, Rodriguez:2016, Berdnikov:2017, Rodriguez:2018}. While the cause of the events is unclear, the accumulation of over a century of photometric observations provides the unique opportunity to search for periodic phenomena on spatial scales of up to $\sim$20 au, similar to the spatial resolution of the ALMA observations. With RW Aurigae continuing to be photometrically monitored from groups like the the American Association of Variable Star Observers (AAVSO) and future ALMA observations of RW Aurigae expected to resolve down to a few AU, we may be able to directly link specific photometric phenomena to observed architectural features. These future observations may provide insight into star-disk interactions and their impact on planet formation.

\subsection{Planet formation inside 10 AU: the disc of DG Tau A observed with the Planet-Earth Building-Blocks Legacy eMERLIN Survey (Emily Drabek-Maunder)}

Planet Earth Building-Blocks -- a Legacy eMERLIN Survey is mapping `pebbles' (cm-sized dust grains) at 5~cm wavelengths for protostars at a range of evolutionary stages and masses. The survey focuses on nearby star-forming regions to systematically study discs with a high potential for planet formation. The 40~mas resolution (5-9 AU) allows us to separate disc zones comparable to where terrestrial and gas giant planets form in our Solar System. The ability to image grain growth within a few AU of young stars is a unique eMERLIN capability, allowing the investigation of how planetary cores are made and the search for protoplanet candidates. 

\begin{figure}
\includegraphics[width=\columnwidth]{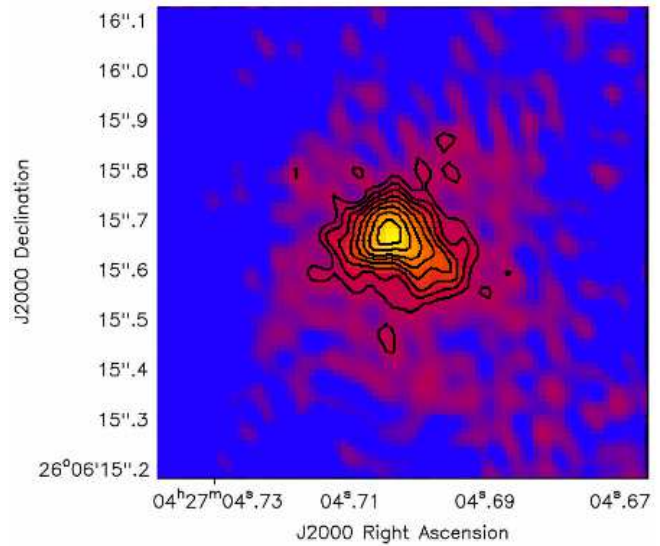}
\caption{eMERLIN $\sim$4.5 cm data of DG Tau A at 3-10~$\sigma$ in 1$\sigma$ increments (with beam size 120$\times$80~mas). The disc major axis is the the SE-NW direction with disc inclined 35 degrees out of the sky plane. The south-western protrusion from the source centre is jet emission which is fully resolved from the disc.\label{fig_drabekmaunder}}
\end{figure}

Commissioning observations of DG Tau A (Class I-II, low-mass) at 4-6 cm detected the resolved disc, easily distinguished from jet emission (Figure~\ref{fig_drabekmaunder}). The extended disc source flux is consistent with an SED predicted by extending the submillimetre and millimetre dust emission down to cm wavelengths (spectral index $\alpha\sim$1.9, consistent with thermal dust emission).  Estimates of the 1 cm dust grain mass indicate the solid mass is very substantial at $\sim$800  $M_{\rm Earth}$, but as expected given that the dust emission is extremely radio bright.  For comparison, the solar system contains only $\sim$44--70 $M_{\rm Earth}$ of solids in the cores of the four gas giants, with $\sim$2 $M_{\rm Earth}$ additionally in the terrestrial planets. 

\subsection{Scaling laws for election cyclotron maser emission: How CMI emission characteristics differ in planetary and stellar magnetospheres (Robert Mutel)}

Electron cyclotron maser emission (ECMI) has been detected both from
planetary and stellar magnetospheres. However, since the plasma environments
for these two classes are quite different, scaling laws used to infer physical properties at the source from observed emission characteristics can also be different. Most importantly, in stellar plasmas electron energies could be much higher than the planetary case. This means that the usual assumption that the observed frequency is equal to the electron gyro-frequency ($\omega_{ce}$), and hence a direct measure of the magnetic field strength, could be significantly in error, because the allowed EMCI emission frequency range (RX-mode) extends down to the relativistic gyrofrequency: $\omega_{ce}/ \gamma < \omega < \omega_{ce}$, where $\gamma$ is the Lorentz factor. This can be significantly less than $\omega_{ce}$ for energetic electrons (e.g., $\gamma=1.5$ for ${\rm E} = 250$~keV). Also, the plasma density, normally assumed to be constrained by the relation $\omega_{pe} \ll \omega_{ce}$, can be much higher, since the cutoff condition is actually

\begin{equation}
\frac{\omega_{pe}}{\omega_{ce}} < \frac{\sqrt{\gamma-1}}{\gamma}
\end{equation}

\begin{figure}
\includegraphics[width=\columnwidth]{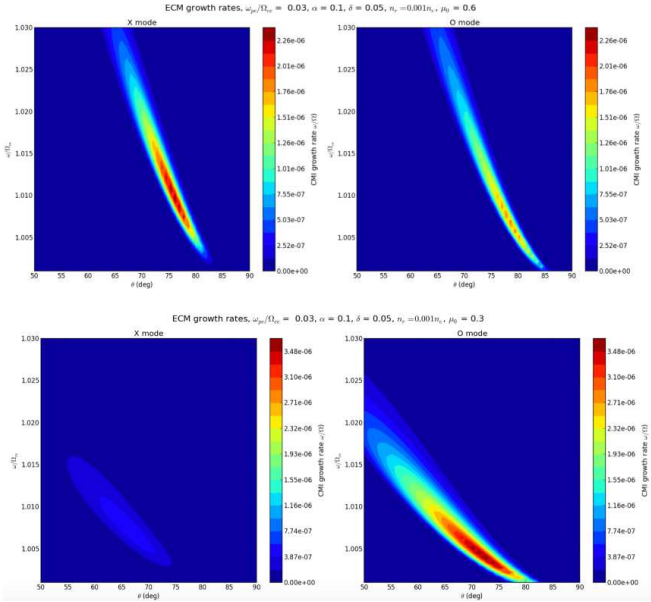}
\caption{Growth rates (color-coded) as a function of beaming angle w.r.t B-
field normal for EMCI emission in R-X and L-O modes, for  $\omega_{pe}/ \omega_{ce}= 0.03$. Upper panel and lower panels have different loss-cone geometries (Mutel et al.~2018). \label{fig_mutel}}
\end{figure}

In addition, the growth rates for L-O mode, which are typically much lower
than R-X mode in planetary environments, may be higher than R-X in a stellar
environment when the density is high. Finally, the beaming angle is strongly
dependent on electron energy and density, resulting in a complex density and
frequency-dependent angular structure, as shown in Figure~\ref{fig_mutel}.

\subsection{Resolving the Radio Surfaces of AGB Stars (Lynn Matthews)}

Stars on the asymptotic giant branch (AGB) return up to 80\% of their
initial mass to the interstellar medium. These stars are one of the primary
sources of dust and chemical enrichment in galaxies.  The mass-losing
winds of AGB stars  are
thought to be primarily dust-driven, but some additional (and still 
poorly understood) physics is required to transport
material from the stellar surface to the wind launch region. This
process likely involves some combination of shocks, pulsation, and/or convection.

Fortuitously, AGB have detectable
``radio photospheres'' lying at roughly twice the classic 
photospheric radius measured at optical wavelengths \citep{rem97}. 
Radio continuum studies of AGB stars therefore
offer a powerful means to probe the atmospheres of these stars just inside the
critical region where dust formation occurs and the stellar wind is launched.
 For the nearest AGB
stars ($d<$150~pc), it is possible to resolve the radio
photosphere at millimeter wavelengths using the long-baseline 
configurations of the VLA and ALMA, enabling the deviation of
fundamental stellar parameters and the imaging of surface features. 

\citet{mat15,mat18} recently observed five
nearby AGB stars with the VLA at  $\lambda\approx$7~mm. All of the stars were
resolved and their radio diameters range from $\sim$5--10~AU
($\sim 4R_{\star}$). 
The photospheric shapes range from nearly round to ellipticities of
$\sim$0.15. However, comparisons with 
observations from several years earlier reveal that the photospheric
parameters (mean size, shape, and/or flux density) have in all cases 
changed with time. 

\begin{figure}
\includegraphics[width=\columnwidth]{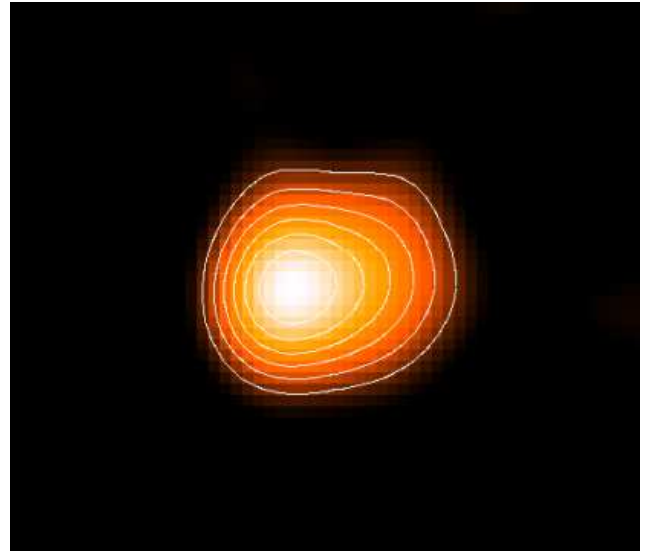}
\caption{The radio photosphere of the AGB star R~Leo, imaged with the VLA at
$\lambda\approx$7~mm. The star exhibits both a non-spherical shape and
a non-uniform surface. This image was produced using
sparse model image reconstruction, enabling
super-resolution of $\sim$0.75 times the dirty beam FWHM of
38~mas. The peak flux density is $\sim$4~mJy beam$^{-1}$ and the image
is $\sim$150~mas on a side. Adapted from \citet{mat18}.
 \label{fig_matthews}}
\end{figure}

Using a novel imaging algorithm
known as sparse model image reconstruction \citep{aki17},
super-resolved images of the VLA-imaged 
stars were derived with effective angular resolutions
of $\sim$20--30~mas 
(\citealp{mat18} and {\it in prep.}; Figure~\ref{fig_matthews}). The resulting images reveal evidence for
irregular shapes and  brightness non-uniformities across the
radio surfaces. Similar features are also indicated in the visibility
domain. These trends are consistent with manifestations of large-scale,
irregular 
convective flows on the stellar surfaces. 

\section{Outlook}

After the end of the structured portion of the splinter session, the participants engaged
in a lively discussion, with the organizers encouraging a focus on future directions for
the field of stellar radio astronomy. Of particular interest was a new class of very large
($\gtrsim$1000~hr) ``X~proposals'' that the U.S. National Radio Astronomy Observatory was
considering offering in future calls. Splinter session co-organizer R.~Osten subsequently led
the preparation and submission of an ``expression of interest'' to the NRAO outlining a
vision for a deep, multi-frequency survey of all of the stars within 10~pc of the Sun. At
the time of the submission of this contribution, it is unknown whether the NRAO will proceed
with the X-proposal program.

\bibliographystyle{cs20proc}
\bibliography{cs20r.bib}

\end{document}